\documentclass[twocolumn,aps,superscriptaddress,nofootinbib]{revtex4-1}
\usepackage[dvips]{graphicx}
\usepackage{latexsym,amssymb,amsmath}
\usepackage{color}
\usepackage[bookmarksnumbered,bookmarksopen,colorlinks,citecolor=red,linkcolor=blue]{hyperref}
\usepackage{mathrsfs}
\usepackage{times}
\usepackage{csquotes}

\begin{document}

\title{Applying Deep Learning Technique to Chiral Magnetic Wave Search}

\author{Yuan-Sheng Zhao}
\affiliation{Physics Department and Center for Particle Physics and Field Theory, Fudan University, Shanghai 200433, China}
\author{Xu-Guang Huang}
\email{huangxuguang@fudan.edu.cn}
\affiliation{Physics Department and Center for Particle Physics and Field Theory, Fudan University, Shanghai 200433, China}
\affiliation{Key Laboratory of Nuclear Physics and Ion-beam Application (MOE), Fudan University, Shanghai 200433, China}
\affiliation{Shanghai Research Center for Theoretical Nuclear Physics, National Natural Science Foundation of China and Fudan University, Shanghai 200438, China}
\date{\today}

\begin{abstract}
The chiral magnetic wave (CMW) is a collective mode in quark-gluon plasma originated from the chiral magnetic effect (CME) and chiral separation effect. Its detection in heavy-ion collisions is challenging due to significant background contamination. In Ref.~\cite{Zhao:2021yjo}, we have constructed a neural network which can accurately identify the CME-related signal from the final-state pion spectra. In this paper, we generalize such a neural network to the case of CMW search. We show that, after a updated training, the neural network can effectively recognize the CMW-related signal. Additionally, we assess the performance of the neural network compared to other known methods for CMW search.\\
\vspace{0pt}
\end{abstract}
\maketitle

\section{Introduction}\label{sec:intro}

The interplay between the chiral anomaly and external electromagnetic or vortical fields can lead to intriguing anomalous transport phenomena in many-body systems with chiral fermions. A notable example is the chiral magnetic effect (CME)~\cite{Kharzeev:2007jp, Fukushima:2008xe}, which induces an electric current aligned with an external magnetic field. In heavy-ion collisions, the CME may cause charge separation relative to the reaction plane, which can potentially be observed by analyzing the azimuthal-angle distribution of charged hadrons using specific observables~\cite{Voloshin:2004vk,Abelev:2009ac}. Other notable anomalous transports include the chiral separation effect (CSE)~\cite{Son:2004tq,Metlitski:2005pr}, the chiral vortical effect~\cite{Erdmenger:2008rm,Banerjee:2008th,Son:2009tf,Landsteiner:2011cp}, and the chiral electric separation effect (CESE)~\cite{Huang:2013iia,Jiang:2014ura}. For reviews, see Refs~\cite{Huang:2015oca,Kharzeev:2015znc,Liu:2020ymh,Kharzeev:2020jxw,Hattori:2022hyo}. 

In the presence of an external magnetic field, the coupled evolution of CME and CSE gives rise to a gapless collective mode known as the chiral magnetic wave (CMW)~\cite{Kharzeev:2010gd}. The CMW can transfer both chirality and electric charge, potentially resulting in distinct charge and chirality distributions. In heavy-ion collisions, given that the fireball contains a small amount of positive charges inherited from the colliding nuclei, theoretical studies have suggested that the CMW can induce a charge quadrupole in the fireball, with an accumulation of positive charges at the tips and negative charges around the equator. As the fireball expands, this quadrupole leads to an imbalance in the elliptic flow of charged pions, specifically, $v_2(\pi^-) > v_2(\pi^+)$~\cite{Burnier:2011bf}. Owing to event-by-event fluctuation of charges, some events could have net negative charges in the fireball thus leading to $v_2(\pi^-) < v_2(\pi^+)$. This characteristic feature of CMW provides a method to detect it in heavy-ion collisions, and a series of experiments have found signals of charged pion elliptic flow consistent with CMW expectations~\cite{Adamczyk:2015eqo,ALICE:2015cjr,CMS:2017pah,STAR:2022zpv,ALICE:2023weh}. However, like CME, CMW in heavy-ion collisions faces strong background noise~\cite{Deng:2012pc,Stephanov:2013tga,Dunlop:2011cf,Bzdak:2013yla,Hatta:2015hca,Xu:2012gf,Hattori:2016emy}, which significantly obscures the observables designed for CMW detection.

In Ref.~\cite{Zhao:2021yjo}, we developed a CME-meter based on convolutional neural network (CNNs) (For reviews of deep learning techniques in nuclear physics, see Refs.~\cite{Boehnlein:2021eym,He:2023urp,He:2023zin,Zhou:2023pti}). After training this CME-meter with AMPT-generated data simulating CME (by introducing initial charge separation into AMPT model~\cite{Lin:2004en}) for Au + Au collisions at 200 GeV, the CME-meter demonstrated exceptional robustness in distinguishing events with CME from those without. Additionally, the CME-meter maintained strong performance across different charge separation fractions, collision energies, and collision systems. This success suggests the potential for creating a similar CMW-meter. As an extension of our earlier work, we aim to increase the upper limit of salience at the cost of some generalization capability. This approach could pave the way for future studies on CMW physics and its detection.

In this paper, we report on the construction and performance of such a CMW-meter. Section II details the training process, including the generation of training samples using AMPT, the structure of the neural network, and the training procedure. Section III examines the analysis of the trained model, including its basic properties, comparisons to flows and observables, and a hypothesis test. Section IV provides a summary of our findings.

\section{Construction and training of the CMW-meter}\label{sec:method}

In this section, we introduce the deep learning model, data set preparation, and training strategies employed in constructing the CMW-meter. The pion spectra of heavy-ion collision final states serve as the input of this deep learning model. Pions carry most of the electric charges in the final state, making them an appropriate representation of charge distribution. A convolutional neural network (CNN) is utilized, trained within a supervised learning scheme to find out the CMW signals. The training data is generated from the string-melting AMPT model~\cite{Lin:2004en}, a transport model which is widely used to simulate the evolution of both partonic and hadronic matter in heavy-ion collisions.

In order to incorporate the CMW effect into the AMPT model, we adopt a global charge quadrupole scheme introduced in Ref.~\cite{Ma:2014iva}. For an AMPT event with $A_{\rm ch}>-0.01$, we interchange the positions of certain $u$ (or $\Bar{d}$) quarks in the initial state with those of $\Bar{u}$ (or $d$) quarks if the former are relatively farther from the reaction plane(RP), while the opposite is done for events with $A_{\rm ch}<-0.01$. Here, $A_{\rm ch}$ stands for the asymmetry of charged particle number, given by $A_{\rm ch} = (N^+-N^-)/(N^++N^-)$, where $N^+$ denotes the number of positively charged particles measured in a given event, and $N^-$ denotes negatively charged particles. The RP of all events is set in the $zOx-$plane. The fraction of particles that are interchanged is represented by a relative percentage with respect to the total number of quarks,
\begin{equation}
    f = \frac{\text{\# Exchanged particles}}{\text{\# All particles}}.
    \label{eq:fcmw}
\end{equation}
According to previous study~\cite{Ma:2014iva}, switching $f=2-3\%$ of quarks generates a CMW signal comparable to experimental observables. For training and validation purposes, we thus choose events with a $f=2\%$ switching fraction. The transition point, $A_{\rm ch}=-0.01$ in this scheme is based on STAR experimental results~\cite{Adamczyk:2015eqo}, where one may find more details. Events at $\sqrt{s_{NN}}=200\,$GeV and different centrality are generated for training and validation. 

There are two primary reasons for training a model that results in bias and overfitting at 200 GeV. Firstly, the pivotal issue pertains to the occurrence of CMW in heavy-ion collisions, rather than the magnitude of the signal. Consequently, any technique that can distinctly distinguish CMW signals from background noise is considered valuable, irrespective of the $\sqrt{s_{NN}}$ or event centrality. Secondly, our research on the application of neural networks for CME detection~\cite{Zhao:2021yjo} confirmed the robustness of the trained network against variations in collision energy and event centrality. The training was successful on the most comprehensive dataset, demonstrating high accuracy levels. This means only small variance in the network's detection feasibility was made by them. Therefore, a model trained on a single energy is capable of enhancing the signal detection in certain events, while still maintaining a considerable degree of generalization. However, further examinations involving various energies and centralities have also been conducted to provide a more nuanced analysis.
\begin{figure}[htbp!]
    \centering
    \includegraphics[width=8cm]{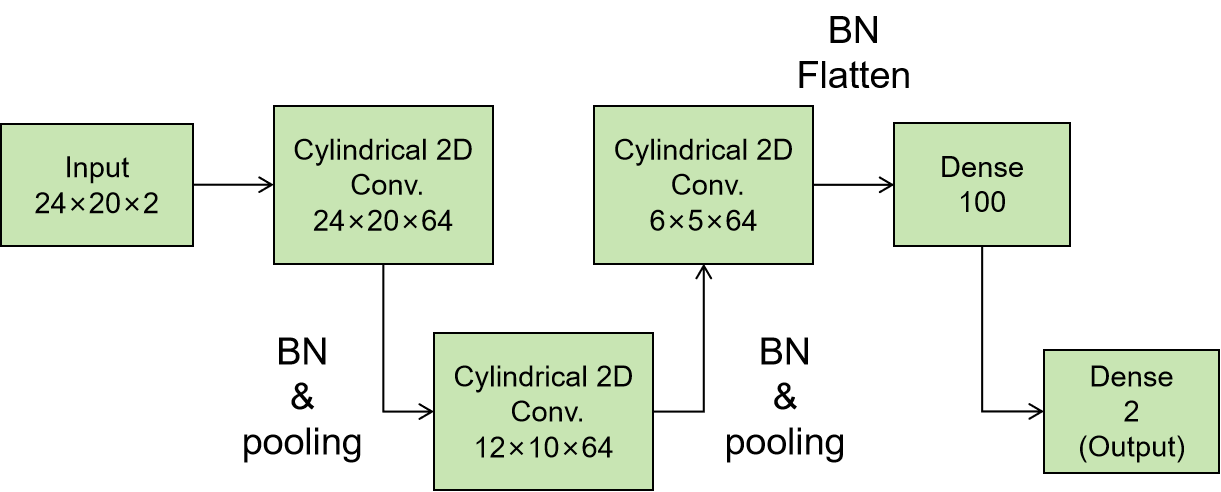}
    \caption{A VGG-like network~\cite{simonyan2015deep} with 4 hidden layers is chosen in this study. Batch normalization(BN) is applied after each hidden layer. The convolutional layers are modified to satisfy the periodic boundary condition of the input data, and each followed by an average pooling. 10\% dropout is set for the second to last dense layers.}
    \label{fig:network}
\end{figure}
The structure of CNN used in this work is shown in Fig.~\ref{fig:network}. There are three 2D-convolutional layers and two dense layer that contain parameters to be fit. Some pooling layers are also applied here, providing proper reduction to data and keeping the network from being too complicated. To encode ``knowledge" about CMW in the model, samples with and without CMW, labeled `1' and `0' separately, are fed to it during training, and the model is set to classify these samples. The last activation function of the network is SoftMax, which returns a pair of numbers $(P_0,\,P_1)$ for this binary classification problem. Samples with $P_1>0.5$ are divided into class `1', the else into class `0'. Therefore, $P_1$ can be interpreted as the probability that a specific sample is recognized by the neural network as containing the CMW signal.  

Data pre-processing involves several steps aimed at converting events into analyzable samples. Initially, the spectra for mid-rapidity ($|\eta|<1$) pions, denoted as $\rho^\pm(p_T, \phi)$, are calculated, with the symbol $\pm$ representing either $\pi^+$ or $\pi^-$, while $p_T$ denotes the transverse momentum in the range of $0-2$ GeV and $\phi$ indicates the azimuthal angle. These spectra are then segmented into histograms consisting of 20 by 24 bins. Second, every spectrum is normalized itself so that the sum of all bins is 1. Subsequently, a random selection of events is made, and for each type of pion, their spectra are averaged bin by bin. These resulting normalized and averaged pion spectra serve as the datasets for the neural network's training, validation, and testing phases. Unless otherwise stated, the number of events we take in the last step is 100 in the rest of this article. The model's training encompasses 250 epochs, within which each epoch contains 64 batches, and each batch comprises 100 samples. In total, 1.6 million samples are generated for the training. 

\section{Performance of the CMW-meter}\label{sec:perf}

\begin{figure}[htbp!]
    \centering
    \includegraphics[width=8cm]{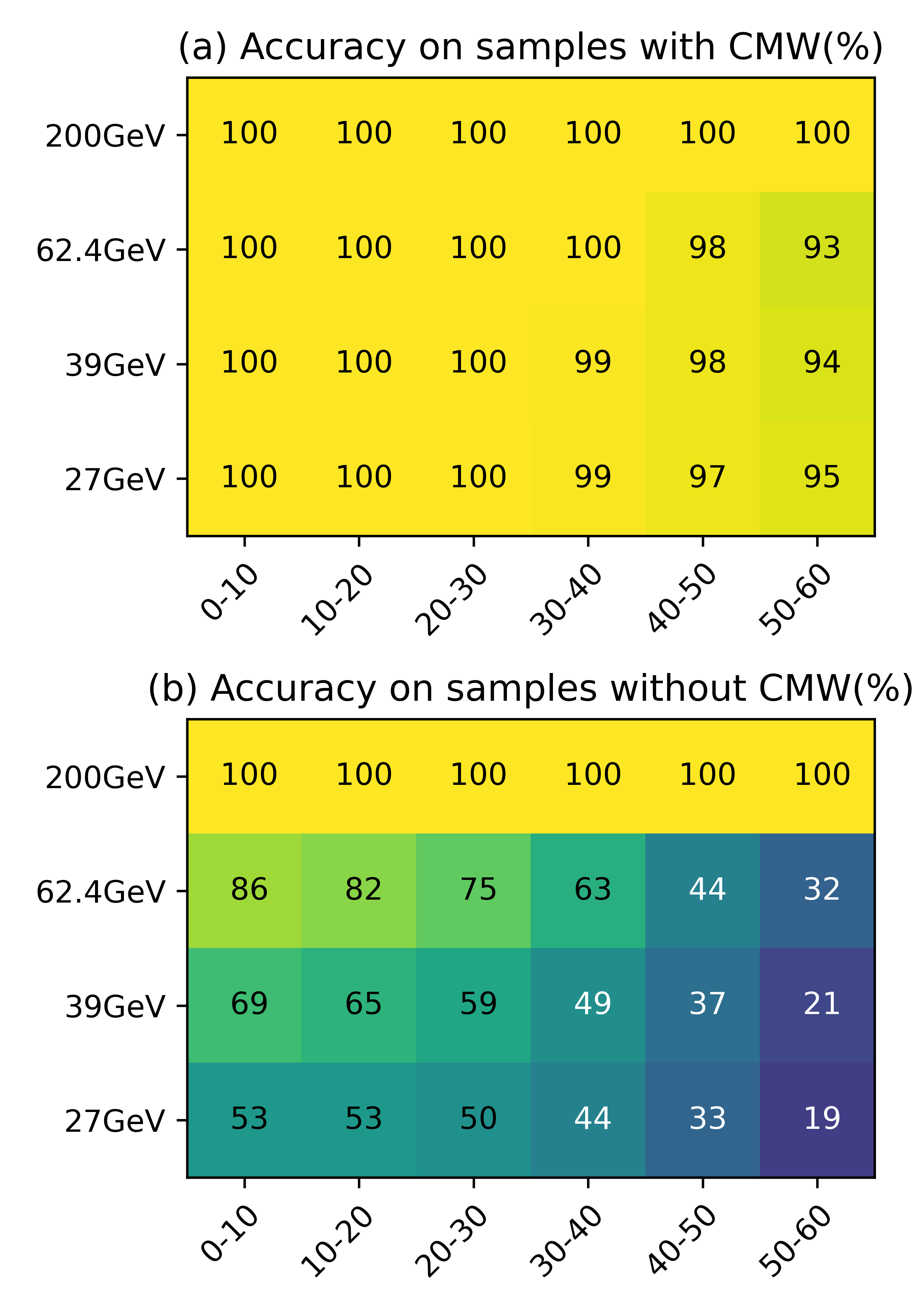}
    \caption{Accuracy of the trained model on samples (a) with CMW, and (b) without CMW. Samples from various $\sqrt{s_{NN}}$ and centralities are considered. The accuracy is remarkably high if the signal is surely encoded in the samples, while those without can be mistaken as containing it, especially at lower energy and more peripheral cases.}
    \label{fig:accuracy}
\end{figure}

\textit{Accuracy, robustness and extrapolations.---} As mentioned above, the model is trained (and also validated) on samples generated at $\sqrt{s_{NN}}=200\,$GeV that mimic final-state CMW behavior. It turns out that the model reaches high accuracy on most events with signal at different $\sqrt{s_{NN}}$ and centrality, as is shown in Fig.~\ref{fig:accuracy}, which indicates preferable generalization of the trained model. Reduction of accuracy can be found at low collision energy and large centrality, of which the reason can be various. Different pattern of CMW, weaker signals, stronger backgrounds or just overfitting, all of them can account for the reduction of accuracy. 
One of the ways to detect CMW in experiments is based on the dependency of charge distribution and flow analysis. Specifically, the linear order dependency of $A_{\rm ch}$ of the difference in charged-particle elliptic flow, 
\begin{equation}
    \Delta v_2 \equiv v_2^- - v_2^+ \simeq r A_{\rm ch},
    \label{eq:deltav2}
\end{equation}
gives a measure for the CMW signal. Here $v_2^-$ and $v_2^+$ are elliptic flows of negative- and positive-charge particles separately, and the slope $r$ is related to the strength of the signal. Experimental results from the STAR experiment for Au + Au collisions~\cite{STAR:2022zpv} indicate that the uncertainty of the $\pi$ slope $r$ increases at lower energies and higher centralities. Although the neural network is trained on pions from a larger kinematic window than used in experimental analyses, suggesting improved completeness and distinguishing capability, its performance aligns with traditional statistical analysis trends ($\Delta v_2 (\pi)$). The decrease in accuracy is likely due to strong backgrounds in those scenarios comparing to the signal strength. However, a new model can be trained using low-energy samples or combined with high-energy samples to create a more comprehensive training set, enhancing robustness. This approach will be for a future, more detailed study. Overall, the trained model's accuracy in decoding the CMW signal is sufficient across all tested $\sqrt{s_{NN}}$ levels, especially when focusing on high-energy samples.

As a potential detector for CMW, a measure of performance is the prediction on non-labeled samples, where accuracy cannot be defined and the sample by sample output becomes important. The two components of the model output are identified as probabilities, \textit{i.e.}, $P_1$ stands for probability that the neural network regards the input spectrum to be with CMW, and $P_0$ is the probability for the other class so the two components satisfy $P_0+P_1=1$. It is obvious that a positive correlation of $P_1$ with CMW signal is expected. Events with different initial charge quadrupole fraction $f$ are simulated, and prepared into samples as mentioned above. Tests on these samples keep the high true-positive accuracy, yet the returned $P_1$ for all $f$ are close to 1 so that little difference is made among them. To make a clear comparison, we enlarge the difference of their output by introducing an additional logit function,
\begin{equation}
    \text{logit}(x)=\text{log}\frac{x}{1-x}\,.
\end{equation}
The logit function is the inverse function of SoftMax, thus acting logit to $P_1$ reveals feature space information that is encoded in the neural network one layer before output. Besides, the logit function is monotonically increasing, so logit($P_1$) keeps the correlation of $P_1$ and $f$ qualitatively. 
\begin{figure}[htbp!]
    \centering
    \includegraphics[width=8.4cm]{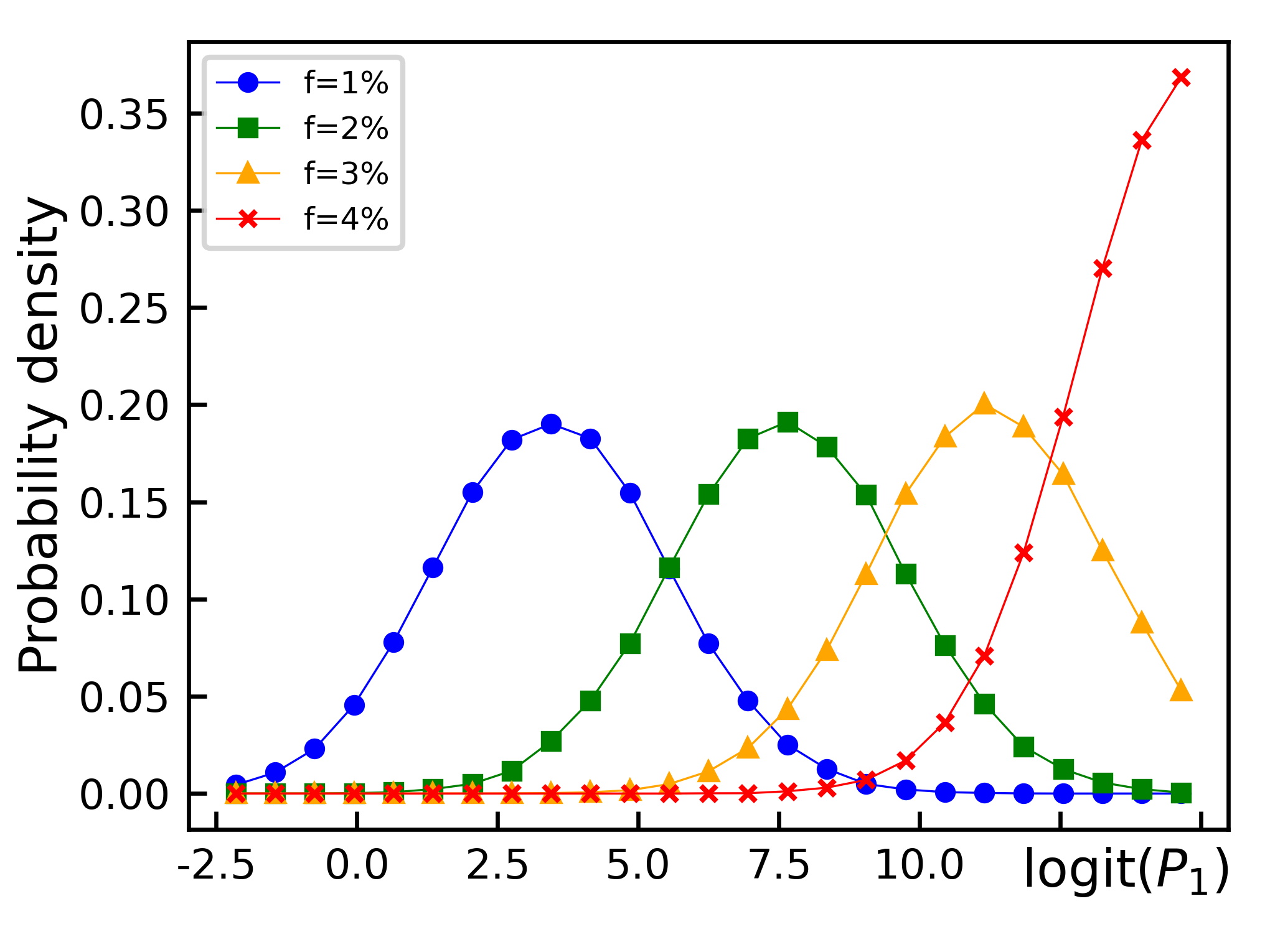}
    \caption{Distribution of logit($P_1$) on events @ $\sqrt{s_{NN}}=200$ GeV, centrality 30-40\%. Tests are done on events with different initial charge quadrupole($f=1,2,3,4\%$). The distribution are normalized to 1.}
    \label{fig:MoreCQ}
\end{figure}

In Fig.~\ref{fig:MoreCQ}, we present the outcomes with varying initial charge quadrupoles. As the $f$ increases, the peak of the logit($P_1$) distribution shifts to the right. In cases where $f$ equals 4\%, $P_1$ approaches 1 so closely that the logit function becomes numerically unstable with single precision calculations. However, the pattern of the $f=4 \%$ distribution is still in line with the general trend. Additionally, the width of the peak remains essentially unaffected by $f$, indicating that the model introduces minimal error and reliably extracts the expected CMW signal. The width of the peak is due to the event-to-event initial-state fluctuations and the method of implementing the initial charge quadrupole. The reasonable extrapolation of $P_1$ for various $f$ values suggests that the CMW strength for $f$ has been correctly aligned to $P_1$ by the neural network. Consequently, it is also indicative of the CMW signal intensity.

The model is also validated in some other tests. In test on no-CMW events generated by UrQMD, it classifies most events correctly into `0' class. To see whether CME signal effects this CMW detector, a test set consists of AMPT events with CME is prepared, and the trained model gives negative prediction mostly. 

\textit{Comparison with observables.---} In above, we demonstrate that the trained model efficiently decodes CMW information from $\rho^{\pm}(p_T, \phi)$, providing a potential measure of CMW in heavy-ion collisions. However, further comparisons with experimental observables are necessary before constructing a measurement based on the model. In Fig.\ref{fig:MoreCQ}, we show that logit($P_1$) is correlated with $f$, which in turn has a positive correlation with the slope $r$. In addition to the slope, the following covariance between $v_n$ and $q_3$, which is essentially a three-particle correlator, is another noteworthy observable~\cite{ALICE:2015cjr},
\begin{equation}
    \lambda_n\equiv\langle v_n q_3\rangle - \langle q_3\rangle\langle v_n\rangle,
\end{equation}
where $v_n$ is the n-th harmonic flow of the event, $q_3$ the charge of the third particle, and $\langle\cdots\rangle$ denotes event average. The differential three-particle correlator, which measures the correlation between the flow at a particular kinematic region, and the charge of the third particle at another particular coordinate, is more convenient when comparing across experiments as no correction for efficiency is needed. In the following we set $n=2$ for correlation with the elliptic flow. Using $v_2^\mp\propto \bar{v}_2\pm r A_{\rm ch}/2$ and $A_{\rm ch}\sim\langle q_3\rangle $, one notice that $\lambda_2\approx\pm r(\langle A_{\rm ch}^2\rangle-\langle A_{\rm ch}\rangle^2)/2$ for positive-charge/negative-charge cases. In the following $\lambda_2$ is obtaiend by making half the difference between the positive-charge and negative-charge cases.
\begin{figure}[htbp!]
    \centering
    \includegraphics[width=8.4cm]{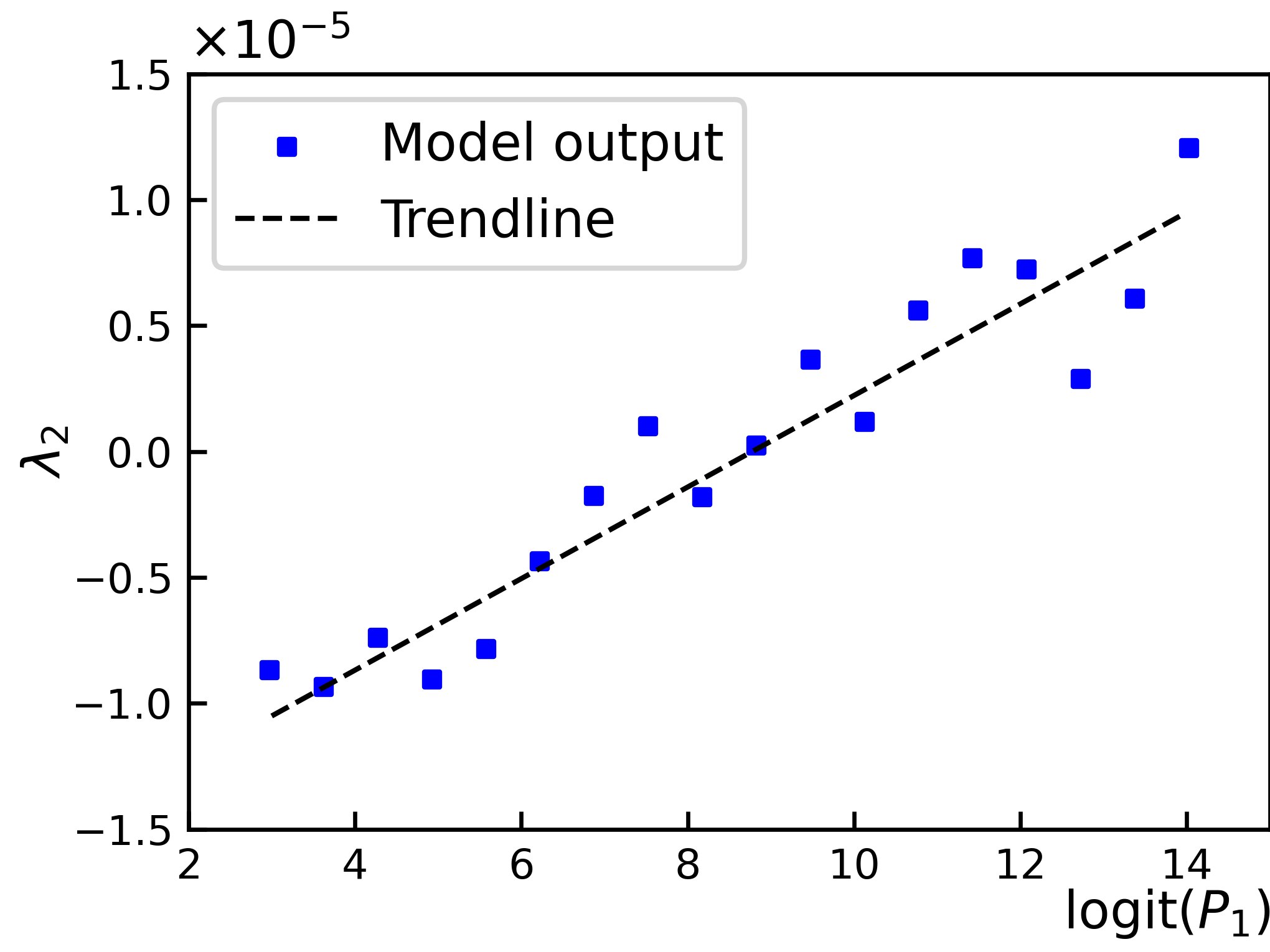}
    \caption{Distribution of logit($P_1$) on events for Au + Au at $\sqrt{s_{NN}}=200$ GeV and centrality 30-40\%. Events are devided into logit($P_1$) bins, and their $\lambda_2$ are averaged separately. Events are all embed with initial charge quadrupole. A range of logit($P_1$) is chosen where most events are included to avoid statistical minority. The three-particle correlator demonstrate clearly a positive correlation with logit($P_1$).}
    \label{fig:3pcorr}
\end{figure}

Figure~\ref{fig:3pcorr} shows the results of the comparison between logit($P_1$) and $\lambda_2$. The average $\lambda_2$ of events increase gently as the response of the model gets stronger. Knowing logit($P_1$) is positvely correlated to CMW signal, it indicates a reasonable trend in $\lambda_2$ when signal gets stronger. This agrees with early studies on $\lambda_2$~\cite{ALICE:2015cjr}, and also shows the model prediction is qualified for measurement. 

\begin{figure}[htbp!]
    \centering
    \includegraphics[width=8.4cm]{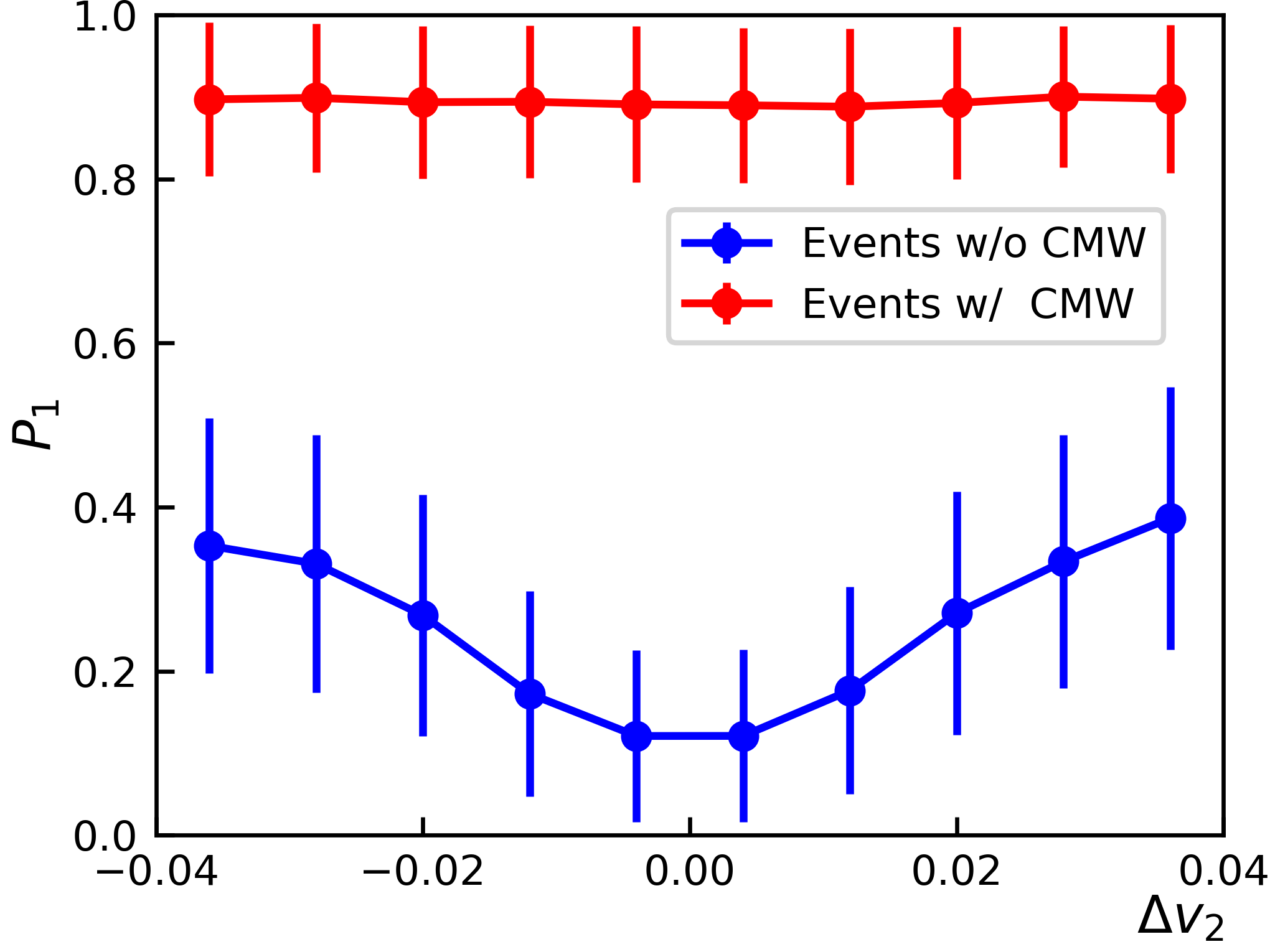}
    \caption{Distribution of $P_1$ on events at $\sqrt{s_{NN}}=200$ GeV against $\Delta v_2$. Events are devided in to 10 $\Delta v_2$ bins, and their $P_1$ are averaged separately. As the magnitude of $\Delta v_2$ increases, the tendency of the model to make a false positive classification also increases. Nevertheless, in events involving an initial quadrupole, the model consistently maintains a high level of accuracy.}
    \label{fig:dv2}
\end{figure}

Performance under backgrounds is necessary before we go further. There are several mechanisms that may cause final-state $\Delta v_2$ and $A_{\rm ch}$ dependency as discussed in Refs.~\cite{Deng:2012pc,Stephanov:2013tga,Dunlop:2011cf,Bzdak:2013yla,Hatta:2015hca,Xu:2012gf,Hattori:2016emy}. To see how the trained neural network works under such backgrounds, the prediction of the neural network with different $\Delta v_2$ range (either with or without initial charge quadrupoles) is studied, and the results are shown in Fig.~\ref{fig:dv2}. For input sample without CMW, the prediction increases when events with larger absolute $\Delta v_2$ are chosen. This shows that the neural network tends to regard events with larger $\Delta v_2$ as events containing CMW signals although they do not actually include CMW signals.  However, it should be emphasized that even in this situation, $P_1$ is still less than $0.5$, meaning that the neural network still correctly classify them as events without CMW. For samples with CMW, the model shows strong robustness against the background, and the model makes classification correctly for all the samples.

\textit{Hypothesis test.---} As previously discussed, the neural network demonstrates good accuracy in predicting the CMW signal and exhibits robustness across different collision energies, centralities, and background effects after training. This makes it feasible to create a CMW-meter based on the neural network. However, the need to average events poses a challenge for deploying this measurement experimentally, as it is not possible to know the charge quadrupole pattern in advance or align events according to their charge distribution patterns. 

However, from a hypothesis test perspective, the CMW-meter also holds experimental feasibility. For a fixed finite number, $M$, of events, one can assume the presence of a sufficiently large residual quadrupole that can be detected through our meter if CMW is believed to exist in these events. Conversely, if no CMW is observed in experiments, the neural network's predictions will consistently fall within the `0' class. As demonstrated in Fig.~\ref{fig:MoreCQ}, the intensity of the CMW signal significantly alters the distribution of logit$(P_1)$ or $P_1$, thus influencing the distribution of $P_1$ itself (denoted by $\text{P}(P_1)$). This distribution responds differently depending on the presence or absence of CMW in the data set. If CMW exists in the heavy-ion collisions, the neural network model's prediction regarding the residual quadrupole of a sample will align with $\text{P}(P_1)$ in the $f \neq 0$ case. In contrast, without any CMW, the distribution will match the $f = 0$ scenario. To establish a reasonable estimation of $\text{P}(P_1)$ for testing $M$ events, we treat $f$ as a latent variable representing CMW in a single event, as defined by the initial charge quadrupole fraction used in this work. Given event-by-event fluctuations, we model $f$ as a random variable following a Gaussian distribution, $f \sim N(\mu, \sigma^2)$, where $\mu$ is the mean of the latent variable $f$, which is expected to be around 0. The variance $\sigma$ is estimated based on~\cite{Ma:2014iva}, where the average of $|f|$ is approximately $2\%$,
\begin{equation}
\begin{aligned}
    2\% =\langle\vert f\vert\rangle &= \int\vert f\vert N^>(\vert f\vert;\sigma^2)\;d\vert f\vert,
    \label{eq: expectation_f}
\end{aligned}
\end{equation}
here $N^>(\vert f\vert;\sigma^2)$ is the half normal distribution, and $\vert f\vert$ is a positive-definite variable because the model prediction is independent to the sign of $f$. Solving (\ref{eq: expectation_f}) gives $\sigma\simeq0.025$. Because we adopt averaged events in preparing the CMW-meter, the way to compose a $\rho(f_\text{eff})$ from single events $\{\rho(f_i)\}$ becomes crucial, where $f_\text{eff}$ is the effective charge quadrupole rate of averaged events. One can choose the arithmetic mean as,
\begin{equation}
    \frac{1}{M}\sum_i^M \rho(f_i) = \rho(\frac{1}{M}\sum_i^M f_i) = \rho(f_\text{eff}).
    \label{eq:feff}
\end{equation}
Therefore, the distribution of $\vert f_\text{eff}\vert$ can be achieved as 
\begin{equation}
 F_\text{eff}\sim N({\mu}/{M}, M\,{\sigma^2}/{M^2})=N(0, {\sigma^2}/{M}),
\end{equation}
with $F_\text{eff}\equiv\vert f_\text{eff}\vert\sim N^>(\sigma^2/M)$. The conditional probability $\text{P}(P_1\vert\;F_\text{eff})$ can be approximated as a \textit{Beta} distribution, 
\begin{equation}
\text{Be}(x;\alpha,\beta)=\frac{\Gamma(\alpha+\beta)}{\Gamma(\alpha)\Gamma(\beta)}x^{\alpha-1}(1-x)^{\beta-1}, 
\end{equation}
with $\alpha$ and $\beta$ the parameters of the \textit{Beta} distribution, and $\Gamma$ is Gamma function. To describe $\text{P}(P_1\vert\;F_\text{eff})$ at any $F_\text{eff}$, we assume $\alpha$ and $\beta$ are functions of $F_\text{eff}$, and fit several sets of $(\alpha, \beta)$ from the fitted beta distribution with polynomial(for $\alpha$) and \textit{Softplus} (for $\beta$, to reach proper asymptotic behavior around $\text{P}_1=1$). 

After parameterizing $\text{P}(P_1\vert\;F_\text{eff})$, $\text{P}(P_1)$ is derived as
\begin{equation}
    \begin{aligned}
    \text{P}(P_1) &= \int\text{P}(P_1\vert\,F_\text{eff})\text{P}(F_\text{eff})\,dF_\text{eff}\\
    &= \int_0^\infty\text{Be}(P_1;\alpha(F_\text{eff}), \beta(F_\text{eff})) N^>(\frac{\sigma^2}{M})\,dF_\text{eff}. 
    \end{aligned}
    \label{final_integral}
\end{equation}

\begin{figure}[htp!]
    \setlength{\belowcaptionskip}{-0.cm}
    \includegraphics[width=8.4cm]{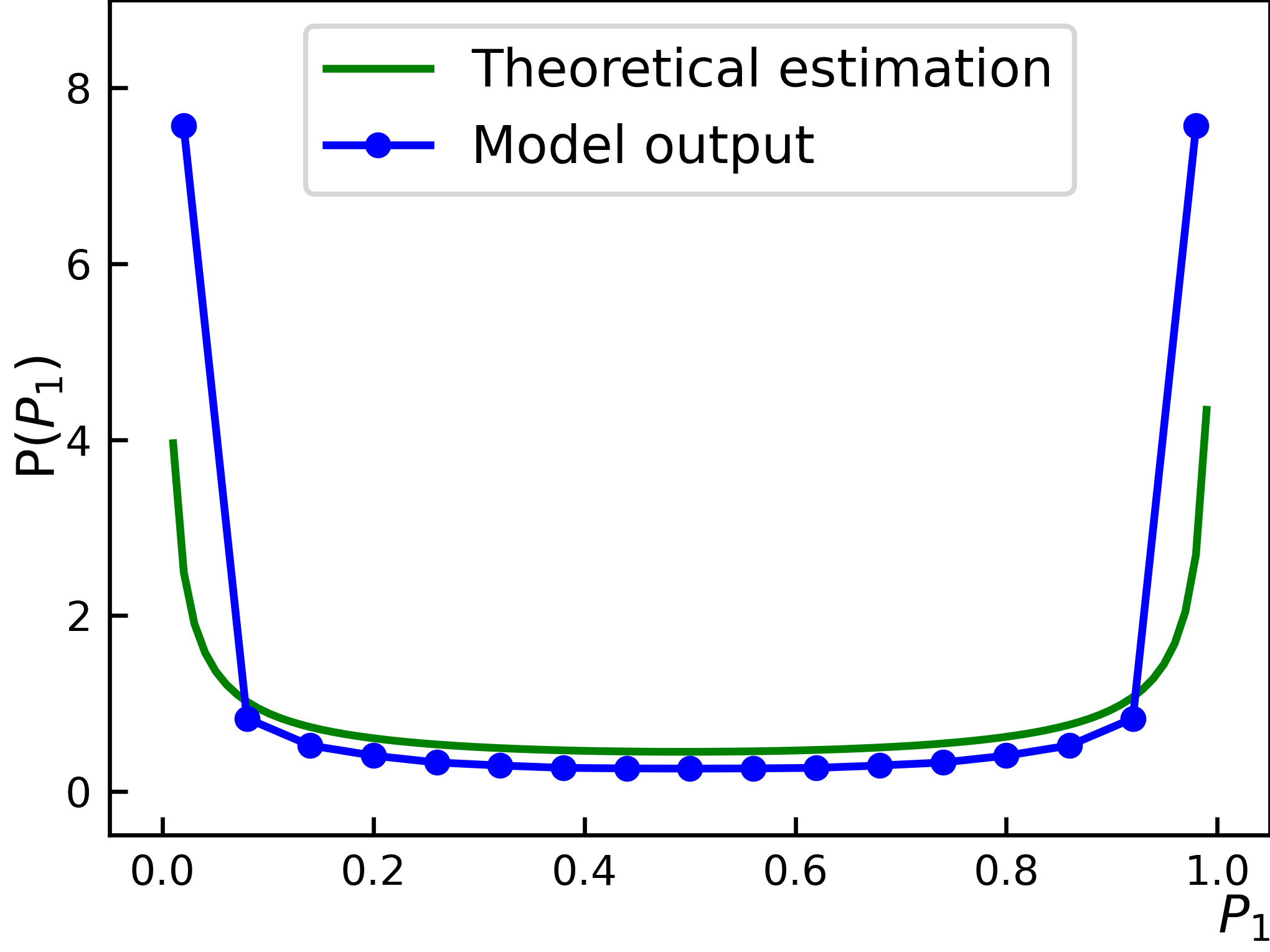}
    \caption{$\text{P}(P_1)$ of AMPT events with $M=25$. The peak of this distribution near $P_1=0$ suggests that the charge quadrupole in most events negate one another, and the peak near $P_1=1$ is because the strong response to some large enough residual quadrupole.}
    \label{fig:criteria}
\end{figure}

The numerical results are shown in Fig.~\ref{fig:criteria}. $\text{P}(P_1)$ of the \enquote{existing CMW} has an obvious rise around $P_1=1$ compared to the \enquote{no CMW} case, which suggests a non-zero probability of composing a large residual quadrupole. With a smaller $M$, the width of $f_\text{eff}$ becomes larger, which allows one to get a visible $P_1$. In Fig.~\ref{fig:criteria} we also present results of randomly mixing events of both charge quadrupole patterns generated by AMPT, where $M=$ 25. For both large- and small-$\text{P}_1$ area, it is consistent with our hypothesis test analysis qualitatively, which indicates that the trained neural network is capable of recognizing charge quadrupole with less averaged events. 

\section{Summary}\label{sec:summ}
In this paper, we propose a deep convolutional neural network model for CMW detection. Building upon our prior work~\cite{Zhao:2021yjo} focused on deep-learning-based CME detection, this model expands its application to CMW detection. We train the neural network using data generated from the AMPT model for Au + Au collisions at 200 GeV, with CMW-like initial charge quadrupole encoded. The trained model exhibits a robust capability to discern events with CMW from those without, and it can quantitatively measure the fraction or strength of the initial charge quadrupole, effectively functioning as a CMW-meter. Furthermore, we validate the model's performance across a broad collision energies and centralities demonstrating its resilience. We have also checked that the trained model can be well qualified even for other collision systems like Zr + Zr and Ru + Ru collisions. Comparative analysis against three-particle correlators and $\Delta v_2$ illustrates the model's effectiveness even in the presence of strong backgrounds. By employing a hypothesis test, an experimentally viable analysis based on the model can be established, wherein the distribution of model predictions serves as an indicator of CMW occurrence in the data.

One drawback of the model is that its brightness is achieved at the cost of generalization ability, as the training data is confined to a narrow range of collision energies. In the future, it would be interesting to enhance the model's generalization capabilities and transform it into an end-to-end CMW meter. 

\textbf{Acknowledgement ---} We acknowledge the useful discussions with L.-X. Wang and K. Zhou. This work is supported by the Natural Science Foundation of Shanghai (Grant No. 23JC1400200), the National Natural Science Foundation of China (Grant No. 12225502, No. 12075061, and  No. 12147101), and the National Key Research and Development Program of China (Grant No. 2022YFA1604900).

\bibliography{reference}

\begin{thebibliography}{39}%
\makeatletter
\providecommand \@ifxundefined [1]{%
 \@ifx{#1\undefined}
}%
\providecommand \@ifnum [1]{%
 \ifnum #1\expandafter \@firstoftwo
 \else \expandafter \@secondoftwo
 \fi
}%
\providecommand \@ifx [1]{%
 \ifx #1\expandafter \@firstoftwo
 \else \expandafter \@secondoftwo
 \fi
}%
\providecommand \natexlab [1]{#1}%
\providecommand \enquote  [1]{``#1''}%
\providecommand \bibnamefont  [1]{#1}%
\providecommand \bibfnamefont [1]{#1}%
\providecommand \citenamefont [1]{#1}%
\providecommand \href@noop [0]{\@secondoftwo}%
\providecommand \href [0]{\begingroup \@sanitize@url \@href}%
\providecommand \@href[1]{\@@startlink{#1}\@@href}%
\providecommand \@@href[1]{\endgroup#1\@@endlink}%
\providecommand \@sanitize@url [0]{\catcode `\\12\catcode `\$12\catcode
  `\&12\catcode `\#12\catcode `\^12\catcode `\_12\catcode `\%12\relax}%
\providecommand \@@startlink[1]{}%
\providecommand \@@endlink[0]{}%
\providecommand \url  [0]{\begingroup\@sanitize@url \@url }%
\providecommand \@url [1]{\endgroup\@href {#1}{\urlprefix }}%
\providecommand \urlprefix  [0]{URL }%
\providecommand \Eprint [0]{\href }%
\providecommand \doibase [0]{http://dx.doi.org/}%
\providecommand \selectlanguage [0]{\@gobble}%
\providecommand \bibinfo  [0]{\@secondoftwo}%
\providecommand \bibfield  [0]{\@secondoftwo}%
\providecommand \translation [1]{[#1]}%
\providecommand \BibitemOpen [0]{}%
\providecommand \bibitemStop [0]{}%
\providecommand \bibitemNoStop [0]{.\EOS\space}%
\providecommand \EOS [0]{\spacefactor3000\relax}%
\providecommand \BibitemShut  [1]{\csname bibitem#1\endcsname}%
\let\auto@bib@innerbib\@empty
\bibitem [{\citenamefont {Zhao}\ \emph {et~al.}(2022)\citenamefont {Zhao},
  \citenamefont {Wang}, \citenamefont {Zhou},\ and\ \citenamefont
  {Huang}}]{Zhao:2021yjo}%
  \BibitemOpen
  \bibfield  {author} {\bibinfo {author} {\bibfnamefont {Y.-S.}\ \bibnamefont
  {Zhao}}, \bibinfo {author} {\bibfnamefont {L.}~\bibnamefont {Wang}}, \bibinfo
  {author} {\bibfnamefont {K.}~\bibnamefont {Zhou}}, \ and\ \bibinfo {author}
  {\bibfnamefont {X.-G.}\ \bibnamefont {Huang}},\ }\href {\doibase
  10.1103/PhysRevC.106.L051901} {\bibfield  {journal} {\bibinfo  {journal}
  {Phys. Rev. C}\ }\textbf {\bibinfo {volume} {106}},\ \bibinfo {pages}
  {L051901} (\bibinfo {year} {2022})},\ \Eprint
  {http://arxiv.org/abs/2105.13761} {arXiv:2105.13761 [hep-ph]} \BibitemShut
  {NoStop}%
\bibitem [{\citenamefont {Kharzeev}\ \emph {et~al.}(2008)\citenamefont
  {Kharzeev}, \citenamefont {McLerran},\ and\ \citenamefont
  {Warringa}}]{Kharzeev:2007jp}%
  \BibitemOpen
  \bibfield  {author} {\bibinfo {author} {\bibfnamefont {D.~E.}\ \bibnamefont
  {Kharzeev}}, \bibinfo {author} {\bibfnamefont {L.~D.}\ \bibnamefont
  {McLerran}}, \ and\ \bibinfo {author} {\bibfnamefont {H.~J.}\ \bibnamefont
  {Warringa}},\ }\href {\doibase 10.1016/j.nuclphysa.2008.02.298} {\bibfield
  {journal} {\bibinfo  {journal} {Nucl. Phys. A}\ }\textbf {\bibinfo {volume}
  {803}},\ \bibinfo {pages} {227} (\bibinfo {year} {2008})},\ \Eprint
  {http://arxiv.org/abs/0711.0950} {arXiv:0711.0950 [hep-ph]} \BibitemShut
  {NoStop}%
\bibitem [{\citenamefont {Fukushima}\ \emph {et~al.}(2008)\citenamefont
  {Fukushima}, \citenamefont {Kharzeev},\ and\ \citenamefont
  {Warringa}}]{Fukushima:2008xe}%
  \BibitemOpen
  \bibfield  {author} {\bibinfo {author} {\bibfnamefont {K.}~\bibnamefont
  {Fukushima}}, \bibinfo {author} {\bibfnamefont {D.~E.}\ \bibnamefont
  {Kharzeev}}, \ and\ \bibinfo {author} {\bibfnamefont {H.~J.}\ \bibnamefont
  {Warringa}},\ }\href {\doibase 10.1103/PhysRevD.78.074033} {\bibfield
  {journal} {\bibinfo  {journal} {Phys. Rev. D}\ }\textbf {\bibinfo {volume}
  {78}},\ \bibinfo {pages} {074033} (\bibinfo {year} {2008})},\ \Eprint
  {http://arxiv.org/abs/0808.3382} {arXiv:0808.3382 [hep-ph]} \BibitemShut
  {NoStop}%
\bibitem [{\citenamefont {Voloshin}(2004)}]{Voloshin:2004vk}%
  \BibitemOpen
  \bibfield  {author} {\bibinfo {author} {\bibfnamefont {S.~A.}\ \bibnamefont
  {Voloshin}},\ }\href {\doibase 10.1103/PhysRevC.70.057901} {\bibfield
  {journal} {\bibinfo  {journal} {Phys. Rev. C}\ }\textbf {\bibinfo {volume}
  {70}},\ \bibinfo {pages} {057901} (\bibinfo {year} {2004})},\ \Eprint
  {http://arxiv.org/abs/hep-ph/0406311} {arXiv:hep-ph/0406311} \BibitemShut
  {NoStop}%
\bibitem [{\citenamefont {Abelev}\ \emph {et~al.}(2009)\citenamefont {Abelev}
  \emph {et~al.}}]{Abelev:2009ac}%
  \BibitemOpen
  \bibfield  {author} {\bibinfo {author} {\bibfnamefont {B.~I.}\ \bibnamefont
  {Abelev}} \emph {et~al.} (\bibinfo {collaboration} {STAR}),\ }\href {\doibase
  10.1103/PhysRevLett.103.251601} {\bibfield  {journal} {\bibinfo  {journal}
  {Phys. Rev. Lett.}\ }\textbf {\bibinfo {volume} {103}},\ \bibinfo {pages}
  {251601} (\bibinfo {year} {2009})},\ \Eprint {http://arxiv.org/abs/0909.1739}
  {arXiv:0909.1739 [nucl-ex]} \BibitemShut {NoStop}%
\bibitem [{\citenamefont {Son}\ and\ \citenamefont
  {Zhitnitsky}(2004)}]{Son:2004tq}%
  \BibitemOpen
  \bibfield  {author} {\bibinfo {author} {\bibfnamefont {D.~T.}\ \bibnamefont
  {Son}}\ and\ \bibinfo {author} {\bibfnamefont {A.~R.}\ \bibnamefont
  {Zhitnitsky}},\ }\href {\doibase 10.1103/PhysRevD.70.074018} {\bibfield
  {journal} {\bibinfo  {journal} {Phys. Rev. D}\ }\textbf {\bibinfo {volume}
  {70}},\ \bibinfo {pages} {074018} (\bibinfo {year} {2004})},\ \Eprint
  {http://arxiv.org/abs/hep-ph/0405216} {arXiv:hep-ph/0405216} \BibitemShut
  {NoStop}%
\bibitem [{\citenamefont {Metlitski}\ and\ \citenamefont
  {Zhitnitsky}(2005)}]{Metlitski:2005pr}%
  \BibitemOpen
  \bibfield  {author} {\bibinfo {author} {\bibfnamefont {M.~A.}\ \bibnamefont
  {Metlitski}}\ and\ \bibinfo {author} {\bibfnamefont {A.~R.}\ \bibnamefont
  {Zhitnitsky}},\ }\href {\doibase 10.1103/PhysRevD.72.045011} {\bibfield
  {journal} {\bibinfo  {journal} {Phys. Rev. D}\ }\textbf {\bibinfo {volume}
  {72}},\ \bibinfo {pages} {045011} (\bibinfo {year} {2005})},\ \Eprint
  {http://arxiv.org/abs/hep-ph/0505072} {arXiv:hep-ph/0505072} \BibitemShut
  {NoStop}%
\bibitem [{\citenamefont {Erdmenger}\ \emph {et~al.}(2009)\citenamefont
  {Erdmenger}, \citenamefont {Haack}, \citenamefont {Kaminski},\ and\
  \citenamefont {Yarom}}]{Erdmenger:2008rm}%
  \BibitemOpen
  \bibfield  {author} {\bibinfo {author} {\bibfnamefont {J.}~\bibnamefont
  {Erdmenger}}, \bibinfo {author} {\bibfnamefont {M.}~\bibnamefont {Haack}},
  \bibinfo {author} {\bibfnamefont {M.}~\bibnamefont {Kaminski}}, \ and\
  \bibinfo {author} {\bibfnamefont {A.}~\bibnamefont {Yarom}},\ }\href
  {\doibase 10.1088/1126-6708/2009/01/055} {\bibfield  {journal} {\bibinfo
  {journal} {JHEP}\ }\textbf {\bibinfo {volume} {01}},\ \bibinfo {pages} {055}
  (\bibinfo {year} {2009})},\ \Eprint {http://arxiv.org/abs/0809.2488}
  {arXiv:0809.2488 [hep-th]} \BibitemShut {NoStop}%
\bibitem [{\citenamefont {Banerjee}\ \emph {et~al.}(2011)\citenamefont
  {Banerjee}, \citenamefont {Bhattacharya}, \citenamefont {Bhattacharyya},
  \citenamefont {Dutta}, \citenamefont {Loganayagam},\ and\ \citenamefont
  {Surowka}}]{Banerjee:2008th}%
  \BibitemOpen
  \bibfield  {author} {\bibinfo {author} {\bibfnamefont {N.}~\bibnamefont
  {Banerjee}}, \bibinfo {author} {\bibfnamefont {J.}~\bibnamefont
  {Bhattacharya}}, \bibinfo {author} {\bibfnamefont {S.}~\bibnamefont
  {Bhattacharyya}}, \bibinfo {author} {\bibfnamefont {S.}~\bibnamefont
  {Dutta}}, \bibinfo {author} {\bibfnamefont {R.}~\bibnamefont {Loganayagam}},
  \ and\ \bibinfo {author} {\bibfnamefont {P.}~\bibnamefont {Surowka}},\ }\href
  {\doibase 10.1007/JHEP01(2011)094} {\bibfield  {journal} {\bibinfo  {journal}
  {JHEP}\ }\textbf {\bibinfo {volume} {01}},\ \bibinfo {pages} {094} (\bibinfo
  {year} {2011})},\ \Eprint {http://arxiv.org/abs/0809.2596} {arXiv:0809.2596
  [hep-th]} \BibitemShut {NoStop}%
\bibitem [{\citenamefont {Son}\ and\ \citenamefont
  {Surowka}(2009)}]{Son:2009tf}%
  \BibitemOpen
  \bibfield  {author} {\bibinfo {author} {\bibfnamefont {D.~T.}\ \bibnamefont
  {Son}}\ and\ \bibinfo {author} {\bibfnamefont {P.}~\bibnamefont {Surowka}},\
  }\href {\doibase 10.1103/PhysRevLett.103.191601} {\bibfield  {journal}
  {\bibinfo  {journal} {Phys. Rev. Lett.}\ }\textbf {\bibinfo {volume} {103}},\
  \bibinfo {pages} {191601} (\bibinfo {year} {2009})},\ \Eprint
  {http://arxiv.org/abs/0906.5044} {arXiv:0906.5044 [hep-th]} \BibitemShut
  {NoStop}%
\bibitem [{\citenamefont {Landsteiner}\ \emph {et~al.}(2011)\citenamefont
  {Landsteiner}, \citenamefont {Megias},\ and\ \citenamefont
  {Pena-Benitez}}]{Landsteiner:2011cp}%
  \BibitemOpen
  \bibfield  {author} {\bibinfo {author} {\bibfnamefont {K.}~\bibnamefont
  {Landsteiner}}, \bibinfo {author} {\bibfnamefont {E.}~\bibnamefont {Megias}},
  \ and\ \bibinfo {author} {\bibfnamefont {F.}~\bibnamefont {Pena-Benitez}},\
  }\href {\doibase 10.1103/PhysRevLett.107.021601} {\bibfield  {journal}
  {\bibinfo  {journal} {Phys. Rev. Lett.}\ }\textbf {\bibinfo {volume} {107}},\
  \bibinfo {pages} {021601} (\bibinfo {year} {2011})},\ \Eprint
  {http://arxiv.org/abs/1103.5006} {arXiv:1103.5006 [hep-ph]} \BibitemShut
  {NoStop}%
\bibitem [{\citenamefont {Huang}\ and\ \citenamefont
  {Liao}(2013)}]{Huang:2013iia}%
  \BibitemOpen
  \bibfield  {author} {\bibinfo {author} {\bibfnamefont {X.-G.}\ \bibnamefont
  {Huang}}\ and\ \bibinfo {author} {\bibfnamefont {J.}~\bibnamefont {Liao}},\
  }\href {\doibase 10.1103/PhysRevLett.110.232302} {\bibfield  {journal}
  {\bibinfo  {journal} {Phys. Rev. Lett.}\ }\textbf {\bibinfo {volume} {110}},\
  \bibinfo {pages} {232302} (\bibinfo {year} {2013})},\ \Eprint
  {http://arxiv.org/abs/1303.7192} {arXiv:1303.7192 [nucl-th]} \BibitemShut
  {NoStop}%
\bibitem [{\citenamefont {Jiang}\ \emph {et~al.}(2015)\citenamefont {Jiang},
  \citenamefont {Huang},\ and\ \citenamefont {Liao}}]{Jiang:2014ura}%
  \BibitemOpen
  \bibfield  {author} {\bibinfo {author} {\bibfnamefont {Y.}~\bibnamefont
  {Jiang}}, \bibinfo {author} {\bibfnamefont {X.-G.}\ \bibnamefont {Huang}}, \
  and\ \bibinfo {author} {\bibfnamefont {J.}~\bibnamefont {Liao}},\ }\href
  {\doibase 10.1103/PhysRevD.91.045001} {\bibfield  {journal} {\bibinfo
  {journal} {Phys. Rev. D}\ }\textbf {\bibinfo {volume} {91}},\ \bibinfo
  {pages} {045001} (\bibinfo {year} {2015})},\ \Eprint
  {http://arxiv.org/abs/1409.6395} {arXiv:1409.6395 [nucl-th]} \BibitemShut
  {NoStop}%
\bibitem [{\citenamefont {Huang}(2016)}]{Huang:2015oca}%
  \BibitemOpen
  \bibfield  {author} {\bibinfo {author} {\bibfnamefont {X.-G.}\ \bibnamefont
  {Huang}},\ }\href {\doibase 10.1088/0034-4885/79/7/076302} {\bibfield
  {journal} {\bibinfo  {journal} {Rept. Prog. Phys.}\ }\textbf {\bibinfo
  {volume} {79}},\ \bibinfo {pages} {076302} (\bibinfo {year} {2016})},\
  \Eprint {http://arxiv.org/abs/1509.04073} {arXiv:1509.04073 [nucl-th]}
  \BibitemShut {NoStop}%
\bibitem [{\citenamefont {Kharzeev}\ \emph {et~al.}(2016)\citenamefont
  {Kharzeev}, \citenamefont {Liao}, \citenamefont {Voloshin},\ and\
  \citenamefont {Wang}}]{Kharzeev:2015znc}%
  \BibitemOpen
  \bibfield  {author} {\bibinfo {author} {\bibfnamefont {D.~E.}\ \bibnamefont
  {Kharzeev}}, \bibinfo {author} {\bibfnamefont {J.}~\bibnamefont {Liao}},
  \bibinfo {author} {\bibfnamefont {S.~A.}\ \bibnamefont {Voloshin}}, \ and\
  \bibinfo {author} {\bibfnamefont {G.}~\bibnamefont {Wang}},\ }\href {\doibase
  10.1016/j.ppnp.2016.01.001} {\bibfield  {journal} {\bibinfo  {journal} {Prog.
  Part. Nucl. Phys.}\ }\textbf {\bibinfo {volume} {88}},\ \bibinfo {pages} {1}
  (\bibinfo {year} {2016})},\ \Eprint {http://arxiv.org/abs/1511.04050}
  {arXiv:1511.04050 [hep-ph]} \BibitemShut {NoStop}%
\bibitem [{\citenamefont {Liu}\ and\ \citenamefont
  {Huang}(2020)}]{Liu:2020ymh}%
  \BibitemOpen
  \bibfield  {author} {\bibinfo {author} {\bibfnamefont {Y.-C.}\ \bibnamefont
  {Liu}}\ and\ \bibinfo {author} {\bibfnamefont {X.-G.}\ \bibnamefont
  {Huang}},\ }\href {\doibase 10.1007/s41365-020-00764-z} {\bibfield  {journal}
  {\bibinfo  {journal} {Nucl. Sci. Tech.}\ }\textbf {\bibinfo {volume} {31}},\
  \bibinfo {pages} {56} (\bibinfo {year} {2020})},\ \Eprint
  {http://arxiv.org/abs/2003.12482} {arXiv:2003.12482 [nucl-th]} \BibitemShut
  {NoStop}%
\bibitem [{\citenamefont {Kharzeev}\ and\ \citenamefont
  {Liao}(2021)}]{Kharzeev:2020jxw}%
  \BibitemOpen
  \bibfield  {author} {\bibinfo {author} {\bibfnamefont {D.~E.}\ \bibnamefont
  {Kharzeev}}\ and\ \bibinfo {author} {\bibfnamefont {J.}~\bibnamefont
  {Liao}},\ }\href {\doibase 10.1038/s42254-020-00254-6} {\bibfield  {journal}
  {\bibinfo  {journal} {Nature Rev. Phys.}\ }\textbf {\bibinfo {volume} {3}},\
  \bibinfo {pages} {55} (\bibinfo {year} {2021})},\ \Eprint
  {http://arxiv.org/abs/2102.06623} {arXiv:2102.06623 [hep-ph]} \BibitemShut
  {NoStop}%
\bibitem [{\citenamefont {Hattori}\ \emph {et~al.}(2022)\citenamefont
  {Hattori}, \citenamefont {Hongo},\ and\ \citenamefont
  {Huang}}]{Hattori:2022hyo}%
  \BibitemOpen
  \bibfield  {author} {\bibinfo {author} {\bibfnamefont {K.}~\bibnamefont
  {Hattori}}, \bibinfo {author} {\bibfnamefont {M.}~\bibnamefont {Hongo}}, \
  and\ \bibinfo {author} {\bibfnamefont {X.-G.}\ \bibnamefont {Huang}},\ }\href
  {\doibase 10.3390/sym14091851} {\bibfield  {journal} {\bibinfo  {journal}
  {Symmetry}\ }\textbf {\bibinfo {volume} {14}},\ \bibinfo {pages} {1851}
  (\bibinfo {year} {2022})},\ \Eprint {http://arxiv.org/abs/2207.12794}
  {arXiv:2207.12794 [hep-th]} \BibitemShut {NoStop}%
\bibitem [{\citenamefont {Kharzeev}\ and\ \citenamefont
  {Yee}(2011)}]{Kharzeev:2010gd}%
  \BibitemOpen
  \bibfield  {author} {\bibinfo {author} {\bibfnamefont {D.~E.}\ \bibnamefont
  {Kharzeev}}\ and\ \bibinfo {author} {\bibfnamefont {H.-U.}\ \bibnamefont
  {Yee}},\ }\href {\doibase 10.1103/PhysRevD.83.085007} {\bibfield  {journal}
  {\bibinfo  {journal} {Phys. Rev. D}\ }\textbf {\bibinfo {volume} {83}},\
  \bibinfo {pages} {085007} (\bibinfo {year} {2011})},\ \Eprint
  {http://arxiv.org/abs/1012.6026} {arXiv:1012.6026 [hep-th]} \BibitemShut
  {NoStop}%
\bibitem [{\citenamefont {Burnier}\ \emph {et~al.}(2011)\citenamefont
  {Burnier}, \citenamefont {Kharzeev}, \citenamefont {Liao},\ and\
  \citenamefont {Yee}}]{Burnier:2011bf}%
  \BibitemOpen
  \bibfield  {author} {\bibinfo {author} {\bibfnamefont {Y.}~\bibnamefont
  {Burnier}}, \bibinfo {author} {\bibfnamefont {D.~E.}\ \bibnamefont
  {Kharzeev}}, \bibinfo {author} {\bibfnamefont {J.}~\bibnamefont {Liao}}, \
  and\ \bibinfo {author} {\bibfnamefont {H.-U.}\ \bibnamefont {Yee}},\ }\href
  {\doibase 10.1103/PhysRevLett.107.052303} {\bibfield  {journal} {\bibinfo
  {journal} {Phys. Rev. Lett.}\ }\textbf {\bibinfo {volume} {107}},\ \bibinfo
  {pages} {052303} (\bibinfo {year} {2011})},\ \Eprint
  {http://arxiv.org/abs/1103.1307} {arXiv:1103.1307 [hep-ph]} \BibitemShut
  {NoStop}%
\bibitem [{\citenamefont {Adamczyk}\ \emph {et~al.}(2015)\citenamefont
  {Adamczyk} \emph {et~al.}}]{Adamczyk:2015eqo}%
  \BibitemOpen
  \bibfield  {author} {\bibinfo {author} {\bibfnamefont {L.}~\bibnamefont
  {Adamczyk}} \emph {et~al.} (\bibinfo {collaboration} {STAR}),\ }\href
  {\doibase 10.1103/PhysRevLett.114.252302} {\bibfield  {journal} {\bibinfo
  {journal} {Phys. Rev. Lett.}\ }\textbf {\bibinfo {volume} {114}},\ \bibinfo
  {pages} {252302} (\bibinfo {year} {2015})},\ \Eprint
  {http://arxiv.org/abs/1504.02175} {arXiv:1504.02175 [nucl-ex]} \BibitemShut
  {NoStop}%
\bibitem [{\citenamefont {Adam}\ \emph {et~al.}(2016)\citenamefont {Adam} \emph
  {et~al.}}]{ALICE:2015cjr}%
  \BibitemOpen
  \bibfield  {author} {\bibinfo {author} {\bibfnamefont {J.}~\bibnamefont
  {Adam}} \emph {et~al.} (\bibinfo {collaboration} {ALICE}),\ }\href {\doibase
  10.1103/PhysRevC.93.044903} {\bibfield  {journal} {\bibinfo  {journal} {Phys.
  Rev. C}\ }\textbf {\bibinfo {volume} {93}},\ \bibinfo {pages} {044903}
  (\bibinfo {year} {2016})},\ \Eprint {http://arxiv.org/abs/1512.05739}
  {arXiv:1512.05739 [nucl-ex]} \BibitemShut {NoStop}%
\bibitem [{\citenamefont {Sirunyan}\ \emph {et~al.}(2019)\citenamefont
  {Sirunyan} \emph {et~al.}}]{CMS:2017pah}%
  \BibitemOpen
  \bibfield  {author} {\bibinfo {author} {\bibfnamefont {A.~M.}\ \bibnamefont
  {Sirunyan}} \emph {et~al.} (\bibinfo {collaboration} {CMS}),\ }\href
  {\doibase 10.1103/PhysRevC.100.064908} {\bibfield  {journal} {\bibinfo
  {journal} {Phys. Rev. C}\ }\textbf {\bibinfo {volume} {100}},\ \bibinfo
  {pages} {064908} (\bibinfo {year} {2019})},\ \Eprint
  {http://arxiv.org/abs/1708.08901} {arXiv:1708.08901 [nucl-ex]} \BibitemShut
  {NoStop}%
\bibitem [{\citenamefont {Abdulhamid}\ \emph {et~al.}(2023)\citenamefont
  {Abdulhamid} \emph {et~al.}}]{STAR:2022zpv}%
  \BibitemOpen
  \bibfield  {author} {\bibinfo {author} {\bibfnamefont {M.~I.}\ \bibnamefont
  {Abdulhamid}} \emph {et~al.} (\bibinfo {collaboration} {STAR}),\ }\href
  {\doibase 10.1103/PhysRevC.108.014908} {\bibfield  {journal} {\bibinfo
  {journal} {Phys. Rev. C}\ }\textbf {\bibinfo {volume} {108}},\ \bibinfo
  {pages} {014908} (\bibinfo {year} {2023})},\ \Eprint
  {http://arxiv.org/abs/2210.14027} {arXiv:2210.14027 [nucl-ex]} \BibitemShut
  {NoStop}%
\bibitem [{\citenamefont {Acharya}\ \emph {et~al.}(2023)\citenamefont {Acharya}
  \emph {et~al.}}]{ALICE:2023weh}%
  \BibitemOpen
  \bibfield  {author} {\bibinfo {author} {\bibfnamefont {S.}~\bibnamefont
  {Acharya}} \emph {et~al.} (\bibinfo {collaboration} {ALICE}),\ }\href
  {\doibase 10.1007/JHEP12(2023)067} {\bibfield  {journal} {\bibinfo  {journal}
  {JHEP}\ }\textbf {\bibinfo {volume} {12}},\ \bibinfo {pages} {067} (\bibinfo
  {year} {2023})},\ \Eprint {http://arxiv.org/abs/2308.16123} {arXiv:2308.16123
  [nucl-ex]} \BibitemShut {NoStop}%
\bibitem [{\citenamefont {Deng}\ and\ \citenamefont
  {Huang}(2012)}]{Deng:2012pc}%
  \BibitemOpen
  \bibfield  {author} {\bibinfo {author} {\bibfnamefont {W.-T.}\ \bibnamefont
  {Deng}}\ and\ \bibinfo {author} {\bibfnamefont {X.-G.}\ \bibnamefont
  {Huang}},\ }\href {\doibase 10.1103/PhysRevC.85.044907} {\bibfield  {journal}
  {\bibinfo  {journal} {Phys. Rev. C}\ }\textbf {\bibinfo {volume} {85}},\
  \bibinfo {pages} {044907} (\bibinfo {year} {2012})},\ \Eprint
  {http://arxiv.org/abs/1201.5108} {arXiv:1201.5108 [nucl-th]} \BibitemShut
  {NoStop}%
\bibitem [{\citenamefont {Stephanov}\ and\ \citenamefont
  {Yee}(2013)}]{Stephanov:2013tga}%
  \BibitemOpen
  \bibfield  {author} {\bibinfo {author} {\bibfnamefont {M.}~\bibnamefont
  {Stephanov}}\ and\ \bibinfo {author} {\bibfnamefont {H.-U.}\ \bibnamefont
  {Yee}},\ }\href {\doibase 10.1103/PhysRevC.88.014908} {\bibfield  {journal}
  {\bibinfo  {journal} {Phys. Rev. C}\ }\textbf {\bibinfo {volume} {88}},\
  \bibinfo {pages} {014908} (\bibinfo {year} {2013})},\ \Eprint
  {http://arxiv.org/abs/1304.6410} {arXiv:1304.6410 [nucl-th]} \BibitemShut
  {NoStop}%
\bibitem [{\citenamefont {Dunlop}\ \emph {et~al.}(2011)\citenamefont {Dunlop},
  \citenamefont {Lisa},\ and\ \citenamefont {Sorensen}}]{Dunlop:2011cf}%
  \BibitemOpen
  \bibfield  {author} {\bibinfo {author} {\bibfnamefont {J.~C.}\ \bibnamefont
  {Dunlop}}, \bibinfo {author} {\bibfnamefont {M.~A.}\ \bibnamefont {Lisa}}, \
  and\ \bibinfo {author} {\bibfnamefont {P.}~\bibnamefont {Sorensen}},\ }\href
  {\doibase 10.1103/PhysRevC.84.044914} {\bibfield  {journal} {\bibinfo
  {journal} {Phys. Rev. C}\ }\textbf {\bibinfo {volume} {84}},\ \bibinfo
  {pages} {044914} (\bibinfo {year} {2011})},\ \Eprint
  {http://arxiv.org/abs/1107.3078} {arXiv:1107.3078 [hep-ph]} \BibitemShut
  {NoStop}%
\bibitem [{\citenamefont {Bzdak}\ and\ \citenamefont
  {Bozek}(2013)}]{Bzdak:2013yla}%
  \BibitemOpen
  \bibfield  {author} {\bibinfo {author} {\bibfnamefont {A.}~\bibnamefont
  {Bzdak}}\ and\ \bibinfo {author} {\bibfnamefont {P.}~\bibnamefont {Bozek}},\
  }\href {\doibase 10.1016/j.physletb.2013.08.003} {\bibfield  {journal}
  {\bibinfo  {journal} {Phys. Lett. B}\ }\textbf {\bibinfo {volume} {726}},\
  \bibinfo {pages} {239} (\bibinfo {year} {2013})},\ \Eprint
  {http://arxiv.org/abs/1303.1138} {arXiv:1303.1138 [nucl-th]} \BibitemShut
  {NoStop}%
\bibitem [{\citenamefont {Hatta}\ \emph {et~al.}(2016)\citenamefont {Hatta},
  \citenamefont {Monnai},\ and\ \citenamefont {Xiao}}]{Hatta:2015hca}%
  \BibitemOpen
  \bibfield  {author} {\bibinfo {author} {\bibfnamefont {Y.}~\bibnamefont
  {Hatta}}, \bibinfo {author} {\bibfnamefont {A.}~\bibnamefont {Monnai}}, \
  and\ \bibinfo {author} {\bibfnamefont {B.-W.}\ \bibnamefont {Xiao}},\ }\href
  {\doibase 10.1016/j.nuclphysa.2015.12.009} {\bibfield  {journal} {\bibinfo
  {journal} {Nucl. Phys. A}\ }\textbf {\bibinfo {volume} {947}},\ \bibinfo
  {pages} {155} (\bibinfo {year} {2016})},\ \Eprint
  {http://arxiv.org/abs/1507.04690} {arXiv:1507.04690 [hep-ph]} \BibitemShut
  {NoStop}%
\bibitem [{\citenamefont {Xu}\ \emph {et~al.}(2012)\citenamefont {Xu},
  \citenamefont {Chen}, \citenamefont {Ko},\ and\ \citenamefont
  {Lin}}]{Xu:2012gf}%
  \BibitemOpen
  \bibfield  {author} {\bibinfo {author} {\bibfnamefont {J.}~\bibnamefont
  {Xu}}, \bibinfo {author} {\bibfnamefont {L.-W.}\ \bibnamefont {Chen}},
  \bibinfo {author} {\bibfnamefont {C.~M.}\ \bibnamefont {Ko}}, \ and\ \bibinfo
  {author} {\bibfnamefont {Z.-W.}\ \bibnamefont {Lin}},\ }\href {\doibase
  10.1103/PhysRevC.85.041901} {\bibfield  {journal} {\bibinfo  {journal} {Phys.
  Rev. C}\ }\textbf {\bibinfo {volume} {85}},\ \bibinfo {pages} {041901}
  (\bibinfo {year} {2012})},\ \Eprint {http://arxiv.org/abs/1201.3391}
  {arXiv:1201.3391 [nucl-th]} \BibitemShut {NoStop}%
\bibitem [{\citenamefont {Hattori}\ and\ \citenamefont
  {Huang}(2017)}]{Hattori:2016emy}%
  \BibitemOpen
  \bibfield  {author} {\bibinfo {author} {\bibfnamefont {K.}~\bibnamefont
  {Hattori}}\ and\ \bibinfo {author} {\bibfnamefont {X.-G.}\ \bibnamefont
  {Huang}},\ }\href {\doibase 10.1007/s41365-016-0178-3} {\bibfield  {journal}
  {\bibinfo  {journal} {Nucl. Sci. Tech.}\ }\textbf {\bibinfo {volume} {28}},\
  \bibinfo {pages} {26} (\bibinfo {year} {2017})},\ \Eprint
  {http://arxiv.org/abs/1609.00747} {arXiv:1609.00747 [nucl-th]} \BibitemShut
  {NoStop}%
\bibitem [{\citenamefont {Boehnlein}\ \emph {et~al.}(2022)\citenamefont
  {Boehnlein} \emph {et~al.}}]{Boehnlein:2021eym}%
  \BibitemOpen
  \bibfield  {author} {\bibinfo {author} {\bibfnamefont {A.}~\bibnamefont
  {Boehnlein}} \emph {et~al.},\ }\href {\doibase 10.1103/RevModPhys.94.031003}
  {\bibfield  {journal} {\bibinfo  {journal} {Rev. Mod. Phys.}\ }\textbf
  {\bibinfo {volume} {94}},\ \bibinfo {pages} {031003} (\bibinfo {year}
  {2022})},\ \Eprint {http://arxiv.org/abs/2112.02309} {arXiv:2112.02309
  [nucl-th]} \BibitemShut {NoStop}%
\bibitem [{\citenamefont {He}\ \emph {et~al.}(2023{\natexlab{a}})\citenamefont
  {He}, \citenamefont {Li}, \citenamefont {Ma}, \citenamefont {Niu},
  \citenamefont {Pei},\ and\ \citenamefont {Zhang}}]{He:2023urp}%
  \BibitemOpen
  \bibfield  {author} {\bibinfo {author} {\bibfnamefont {W.}~\bibnamefont
  {He}}, \bibinfo {author} {\bibfnamefont {Q.}~\bibnamefont {Li}}, \bibinfo
  {author} {\bibfnamefont {Y.}~\bibnamefont {Ma}}, \bibinfo {author}
  {\bibfnamefont {Z.}~\bibnamefont {Niu}}, \bibinfo {author} {\bibfnamefont
  {J.}~\bibnamefont {Pei}}, \ and\ \bibinfo {author} {\bibfnamefont
  {Y.}~\bibnamefont {Zhang}},\ }\href {\doibase 10.1007/s11433-023-2116-0}
  {\bibfield  {journal} {\bibinfo  {journal} {Sci. China Phys. Mech. Astron.}\
  }\textbf {\bibinfo {volume} {66}},\ \bibinfo {pages} {282001} (\bibinfo
  {year} {2023}{\natexlab{a}})},\ \Eprint {http://arxiv.org/abs/2301.06396}
  {arXiv:2301.06396 [nucl-th]} \BibitemShut {NoStop}%
\bibitem [{\citenamefont {He}\ \emph {et~al.}(2023{\natexlab{b}})\citenamefont
  {He}, \citenamefont {Ma}, \citenamefont {Pang}, \citenamefont {Song},\ and\
  \citenamefont {Zhou}}]{He:2023zin}%
  \BibitemOpen
  \bibfield  {author} {\bibinfo {author} {\bibfnamefont {W.-B.}\ \bibnamefont
  {He}}, \bibinfo {author} {\bibfnamefont {Y.-G.}\ \bibnamefont {Ma}}, \bibinfo
  {author} {\bibfnamefont {L.-G.}\ \bibnamefont {Pang}}, \bibinfo {author}
  {\bibfnamefont {H.-C.}\ \bibnamefont {Song}}, \ and\ \bibinfo {author}
  {\bibfnamefont {K.}~\bibnamefont {Zhou}},\ }\href {\doibase
  10.1007/s41365-023-01233-z} {\bibfield  {journal} {\bibinfo  {journal} {Nucl.
  Sci. Tech.}\ }\textbf {\bibinfo {volume} {34}},\ \bibinfo {pages} {88}
  (\bibinfo {year} {2023}{\natexlab{b}})},\ \Eprint
  {http://arxiv.org/abs/2303.06752} {arXiv:2303.06752 [hep-ph]} \BibitemShut
  {NoStop}%
\bibitem [{\citenamefont {Zhou}\ \emph {et~al.}(2024)\citenamefont {Zhou},
  \citenamefont {Wang}, \citenamefont {Pang},\ and\ \citenamefont
  {Shi}}]{Zhou:2023pti}%
  \BibitemOpen
  \bibfield  {author} {\bibinfo {author} {\bibfnamefont {K.}~\bibnamefont
  {Zhou}}, \bibinfo {author} {\bibfnamefont {L.}~\bibnamefont {Wang}}, \bibinfo
  {author} {\bibfnamefont {L.-G.}\ \bibnamefont {Pang}}, \ and\ \bibinfo
  {author} {\bibfnamefont {S.}~\bibnamefont {Shi}},\ }\href {\doibase
  10.1016/j.ppnp.2023.104084} {\bibfield  {journal} {\bibinfo  {journal} {Prog.
  Part. Nucl. Phys.}\ }\textbf {\bibinfo {volume} {135}},\ \bibinfo {pages}
  {104084} (\bibinfo {year} {2024})},\ \Eprint
  {http://arxiv.org/abs/2303.15136} {arXiv:2303.15136 [hep-ph]} \BibitemShut
  {NoStop}%
\bibitem [{\citenamefont {Lin}\ \emph {et~al.}(2005)\citenamefont {Lin},
  \citenamefont {Ko}, \citenamefont {Li}, \citenamefont {Zhang},\ and\
  \citenamefont {Pal}}]{Lin:2004en}%
  \BibitemOpen
  \bibfield  {author} {\bibinfo {author} {\bibfnamefont {Z.-W.}\ \bibnamefont
  {Lin}}, \bibinfo {author} {\bibfnamefont {C.~M.}\ \bibnamefont {Ko}},
  \bibinfo {author} {\bibfnamefont {B.-A.}\ \bibnamefont {Li}}, \bibinfo
  {author} {\bibfnamefont {B.}~\bibnamefont {Zhang}}, \ and\ \bibinfo {author}
  {\bibfnamefont {S.}~\bibnamefont {Pal}},\ }\href {\doibase
  10.1103/PhysRevC.72.064901} {\bibfield  {journal} {\bibinfo  {journal} {Phys.
  Rev. C}\ }\textbf {\bibinfo {volume} {72}},\ \bibinfo {pages} {064901}
  (\bibinfo {year} {2005})},\ \Eprint {http://arxiv.org/abs/nucl-th/0411110}
  {arXiv:nucl-th/0411110} \BibitemShut {NoStop}%
\bibitem [{\citenamefont {Ma}(2014)}]{Ma:2014iva}%
  \BibitemOpen
  \bibfield  {author} {\bibinfo {author} {\bibfnamefont {G.-L.}\ \bibnamefont
  {Ma}},\ }\href {\doibase 10.1016/j.physletb.2014.06.074} {\bibfield
  {journal} {\bibinfo  {journal} {Phys. Lett. B}\ }\textbf {\bibinfo {volume}
  {735}},\ \bibinfo {pages} {383} (\bibinfo {year} {2014})},\ \Eprint
  {http://arxiv.org/abs/1401.6502} {arXiv:1401.6502 [nucl-th]} \BibitemShut
  {NoStop}%
\bibitem [{\citenamefont {Simonyan}\ and\ \citenamefont
  {Zisserman}(2015)}]{simonyan2015deep}%
  \BibitemOpen
  \bibfield  {author} {\bibinfo {author} {\bibfnamefont {K.}~\bibnamefont
  {Simonyan}}\ and\ \bibinfo {author} {\bibfnamefont {A.}~\bibnamefont
  {Zisserman}},\ }\href@noop {} {\enquote {\bibinfo {title} {Very deep
  convolutional networks for large-scale image recognition},}\ } (\bibinfo
  {year} {2015}),\ \Eprint {http://arxiv.org/abs/1409.1556} {arXiv:1409.1556
  [cs.CV]} \BibitemShut {NoStop}%
\end{thebibliography}%

\end{document}